\documentclass{PoS}
\usepackage{amsmath}
\usepackage{graphicx}
\usepackage{amssymb}
\usepackage{comment}
\usepackage{cite}
\numberwithin{equation}{section}

\newcommand{\mynote}[1]{}

\newlength{\dummysp}
\newcommand{\tr}{\mathop{{\hbox{Tr} \, }}\nolimits}

\newcommand{\beq}{\begin{eqnarray}}
\newcommand{\eeq}{\end{eqnarray}}
\newcommand{\nnn}{ \nonumber \\ }
\newcommand{\ddd}{ \nnn && }

\newcommand{\vev}[1]{{\langle #1 \rangle}}

\newcommand{\ord}[1]{{{\cal O}(#1)}}
\newcommand{\gappeq}{\mathrel{\rlap {\raise.5ex\hbox{$>$}}
{\lower.5ex\hbox{$\sim$}}}}
\newcommand{\lappeq}{\mathrel{\rlap{\raise.5ex\hbox{$<$}}
{\lower.5ex\hbox{$\sim$}}}}
\newcommand{\myref}[1]{(\ref{#1})}

\newcommand{\ben}{\begin{enumerate}}
\newcommand{\een}{\end{enumerate}}
\newcommand{\bit}{\begin{itemize}}
\newcommand{\eit}{\end{itemize}}
\newcommand{\Cbf}{{\bf C}}

\newcommand{\Ncal}{{\cal N}}

\newcommand{\Ocal}{{\cal O}}

\newcommand{\Zbf}{{\bf Z}}

\newcommand{\cQ}{{\cal Q}}
\newcommand{\cU}{{\cal U}}

\newcommand{\cUb}{{\overline{\cal U}}}

\def\[{\left [}
\def\]{\right ]}
\def\({\left (}
\def\){\right )}

\title{S-duality in lattice super Yang-Mills}

\ShortTitle{S-duality}

\author{\speaker{Joel Giedt}\thanks{JG was partially supported by the Department of Energy grant DE-SC0013496.
        SC and DS were partially supported by Department of Energy grant DE-SC0009998.}\\
        Department of Physics, Applied Physics and Astronomy, Rensselaer Polytechnic Institute, Troy, NY USA\\
        E-mail: \email{giedtj@rpi.edu}}

\author{Simon Catterall\\
        Department of Physics, Syracuse University, Syracuse, NY USA\\
        E-mail: \email{smcatterall@gmail.com}}

\author{Poul Damgaard\\
        Neils Bohr Institute, Copenhagen, Denmark\\
        E-mail: \email{phdamg@nbi.dk}}

\author{David Schaich\\
        Institute for Theoretical Physics, University of Bern, Bern, Switzerland\\
        E-mail: \email{daschaich@gmail.com}}

\abstract{We present a progress report on studying
          S-duality in lattice $\Ncal=4$ super Yang-Mills.
This is being done through a computation of $1/2$-BPS states on the Coulomb branch,
especially the 't Hooft--Polyakov monopole and the W boson.  Key to these
calculations is the use of twisted and C-periodic boundary conditions.
  In addition we describe a variational method to disentangle operators
with definite scaling dimension, particularly the Konishi
and supergravity operators.}

\FullConference{34th annual International Symposium on Lattice Field Theory\\
		24-30 July 2016\\
		University of Southampton, UK}

\begin{document}

\section{Introduction}
We report on certain aspects of our continuing theoretical studies
of $\Ncal=4$ super Yang-Mills (SYM)
using lattice gauge theory techniques.
There are many reasons for formulating and studying such theories
using this first principles approach, with the goal in mind
of repeating the successes of lattice quantum chromodynamics---which
are quite substantial.  Continuum tools such as nonrenormalization
theorems, holomorphy, anomaly matching and the computation of
BPS protected quantities are quite powerful and have allowed
for impressive progress in understanding supersymmetric field
theories over several decades.  However, there remain many unanswered
questions.  These include the nonperturbative spectrum in strongly
coupled gauge theories,\footnote{On the Coulomb
branch of $\Ncal=4$ SYM, the theory is gapped and
there will be a particle spectrum.  In fact the
BPS saturated states are quite interesting to us
in testing S-duality, and will be described below.
One of the chief goals of our research is to verify a
continuum formula for the spectrum of particles
that is supposed to be exact---Eq.~\myref{bps}.} 
holographic duality for quantities that
are not BPS protected, renormalization of nonholomorphic
quantites such as the K\"ahler potential, and many other aspects
of these theories that need to be studied 
at a more detailed, quantitative level.

Dualities are useful because this understanding of seemingly
different theories in fact unifies them under an umbrella of equivalent
descriptions.  In many examples there is a self-duality, where the
dual theory has an identical action except that the parameters
are transformed; the self-dual point for parameters is
often associated with a critical point, as in the Kramers-Wannier
duality ($T \sim 1/T$, with $T$ temperature) 
of the two-dimensional (classical) Ising model.  
Also, supersymmetric gauge theories have a rich vacuum
structure because of the extension of spacetime symmetries; complexities of the
vacuum are much better understood by studying the implications
of dualities, such as has been done in \cite{Aharony:2013hda}.  Global aspects
of gauge theories, such as consistency constraints on line operators,
are very well addressed by a concrete lattice formulation such as we describe
here.  Our current studies focus on a strongly coupled field theory
where many exact results are available---an essential aspect since
formulating supersymmetric systems on the lattice is a difficult
problem.\footnote{Lattice discretization necessarily breaks supersymmetry
at the scale of the lattice spacing because the full supersymmetry
algebra closes on the generators of infinitesmal spacetime translation.
To recover supersymmetry in the continuum limit, it must emerge
as a symmetry as long distances, just like Lorentz invariance.}

\section{Status of lattice $\Ncal=4$ SYM}
Over the last few years we have been studying a lattice 
formulation of $\Ncal=4$ SYM that is based
on a particular (Marcus) topological twist \cite{Marcus:1995mq}
of the continuum theory \cite{Catterall:2007kn}, and which is equivalent 
to formulating the theory through orbifolding a matrix model \cite{Kaplan:2005ta}.  
The covariant derivatives are chosen
in such a way that spectral doubling is avoided, based on old works involving the formulation of
K\"ahler-Dirac fermions on the lattice \cite{Rabin:1981qj,Becher:1981cb,Becher:1982ud,Banks:1982iq}.
All of the numerical tests that we have performed on
the theory show that there is no sign problem for the fermion measure (which is a pfaffian
in this case, rather than a determinant), provided we use antiperiodic
boundary conditions for the fermions and the 't Hooft coupling is not too large \cite{Catterall:2014vka}.
 
We have studied the renormalization of the theory
and have shown that no new relevant or marginal operators are generated in the flow to long
distances \cite{Catterall:2011pd}.  Instead, the coefficients of the terms 
that are already in the action will be
modified from their tree-level values.  (To avoid one new relevant operator, we invoke our
results regarding moduli space not being lifted at any finite order 
in perturbation theory \cite{Catterall:2011pd},
and the assumption that this will also be the case nonperturbatively \cite{Catterall:2014mha}.)
We have found that the $\beta$ functions for these coefficients all vanish
at one loop in lattice perturbation theory, which shows that the lattice
theory is equivalent to the continuum theory at one loop.  
These coefficients
all experience a logarithmic flow at higher orders because they correspond to marginal operators near the
Gaussian fixed point.  All of the flow is due to lattice artifacts; after taking account
of the ability to redefine the fields, we find \cite{Catterall:2014mha} that 
the fine-tuning necessary to restore
the full $\Ncal=4$ supersymmetry consists of tuning to a submanifold in the space of
couplings with codimension 1.  In this sense, the fine-tuning is no worse than for
Wilson fermions in lattice QCD.  This is to be compared to formulating
$\Ncal=4$ SYM with Wilson fermions, which would require eight
parameters to be fine-tuned \cite{Catterall:2014mha}. 

We are continuing our studies of this theory
by using Monte Carlo renormalization group techniques to identify the correct tuning
as a function of the lattice spacing.  Initial studies of this
were performed for Wilson loops in \cite{Catterall:2014mha}.  We found that
the Wilson loops on a blocked fine lattice and a coarse lattice agreed
with each other within errors without any fine-tuning being
necessary.\footnote{Optimization of a blocking parameter was utilized.}
We view this as a reflection of the fact that the
coefficients are running very slowly due to the approximate
supersymmetry of the lattice action.  Further studies are underway
using operators constructed out of fermions, which should provide
tests that are in some sense ``orthogonal'' to or ``independent'' of those involving Wilson loops.

We have found in recent work \cite{Catterall:2013roa} that
any fine-tuning that restores a discrete version of the $SU(4)_R$ global
symmetry of $\Ncal=4$ will automatically recover the full 16-supercharge
supersymmetry.\footnote{This is a consequence of the exact scalar supercharge
and enhanced point group symmetry that we preserve on the $A_4^*$ lattice.
A hypercubic lattice would not have this nice property.}
Thus we are able to test the amount of supersymmetry breaking by measuring the
difference between $n \times n$ Wilson loops\footnote{As described below,
our link variables include scalars, and therefore transform nontrivially
under the $R$ symmetry.} and the corresponding loops rotated by
the discrete $R$ transformation.
The violation of the $R$ symmetry is $\ord{10}$ per
cent \cite{Catterall:2014vka}, and we expect that this will also be the case for the
other 15 supercharges $\cQ_a, \cQ_{ab}$ ($a,b=1,\ldots,5$) that are not scalars in the twisted
formulation.  This indicates that there will ultimately be a need for
some fine-tuning in order to recover the desired continuum theory.

\section{S-duality}
One of our current efforts is to study a key feature of
$\Ncal=4$ SYM, namely S-duality.  
We will do this by measuring part of the spectrum
of $\frac{1}{2}$-BPS states, especially the W-boson and 't Hooft--Polyakov magnetic
monopole on the Coulomb branch.
This spectrum is given by
\beq
M_{p,q}=vg|p+q\tau|=vg\sqrt{\left(p+\frac{\theta}{2\pi}q\right)^2+\left(\frac{4 \pi q}{g^2}\right)^2}
\label{bps}
\eeq
Here $p$ is the electric charge, $q$ is the magnetic charge, $v$ is the scalar
field expectation value that spontaneously breaks the gauge symmetry on the Coulomb
branch of $\Ncal=4$ SYM moduli space, and $\tau$ is the complexified coupling:
$\tau = (\theta/2\pi) + i (4 \pi/g^2)$.
In this last expression $g$ is the gauge coupling and $\theta$ is the parameter
that described the partition function in terms of topological sectors
$Z = \sum_\nu e^{i \nu \theta} Z_\nu$.
The index $\nu$ is the topological charge of a particular sector.  The full duality
group for $SU(N)$ gauge theory (which is simply-laced) is $SL(2,Z)$.  This
acts on the charges of the $\frac{1}{2}$-BPS states (which are generically dyons)
according to ($a,b,c,d \in \Zbf$)
\beq
\binom{p}{q} \to \binom{p'}{q'} = \begin{pmatrix} a & b \cr c & d \end{pmatrix} \binom{p}{q} = \binom{a p + b q}{c p + d q}, \quad ad-bc=1
\eeq
while for the complexified coupling a projective transformation is made:
$\tau \to \tau' = (a \tau + b)/(c \tau + d)$.
Verifying all of these features in our numerical simulations,
to the extent that it is possible,\footnote{Unfortunately, the simulations
must be restricted to $\theta=0$ to avoid a sign problem.}
will provide a nonperturbative check on S-duality
using a first principles approach.  In particular, the W boson $M_{1,0}$ and
the 't Hooft--Polyakov monopole $M_{0,1}$ are mapped into each other under
\beq
S = \begin{pmatrix} 0 & -1 \cr 1 & 0 \end{pmatrix}
\eeq
so we will measure both of these states as a function of the coupling $g$.
Furthermore, because the spectrum in \myref{bps} is BPS, it is an exact
prediction which we can aim to verify numerically.  Doing so will give
further confidence in both the lattice techniques and the continuum arguments.

In order to perform this study, we have to push the lattice theory out onto the Coulomb
branch where $U(N) \to U(1)^N$ spontaneous
gauge symmetry breaking occurs.  We do this by adding a small negative mass-squared for one of the scalars
and then remove it in the thermodynamic limit.  The form of that mass is
\beq
\Delta S = -F \sum_x {\rm Tr} P^2(x),
\quad P = \( {\cal U}_m^\dagger {\cal U}_m \)_\text{traceless},
\quad {\rm n.s.}m
\eeq
where $\cU_m$, $m=1,\ldots,5$ are the $GL(N,\Cbf)$ valued link fields,
due to complexification of the gauge field (this, together 
with the 4d $\to$ 5d lift, is how scalars are
incorporated in the twisted/orbifold approach).  The mass term can be seen
by considering the continuum limit of this expression, which is determined by
the following expansion of the link fields:
\beq
{\cal U}_m = \frac{1}{a} + {\cal A}_m, \quad {\cal A}_m = A_m + i B_m
\label{Uexpansion}
\eeq
Here, $A_m = A_m^i t^i$ corresponds to the gauge field (up to a subtlety regarding
the sixth scalar\footnote{To be precise, the sixth scalar is
given by $\phi_6 = (1/\sqrt{5}) \sum_{m=1}^5 A_m$ for the $A_4^*$
lattice that we are using in our formulation.}), 
$B_m = B_m^i t^i$ are scalars, and we use anti-Hermitian
generators $t^i$ of $U(N)$.  In order to recover the theory with the continuum symmetries, it is necessary to
remove the mass term in the thermodynamic limit, $F \sim 1/V$, where $V$
is the spacetime volume (we use a 4d torus in our lattice formulation).

The mass of the 't Hooft--Polyakov monopole has been computed on the lattice
previously in the simpler Georgi-Glashow 
model \cite{Davis:2000kv,Davis:2001mg,Rajantie:2005hi}.  One computes the free energy
difference between partition functions with twisted boundary conditions\footnote{Here
we specify to the gauge group $SU(2)$ for purposes of illustration; it is known how
to generalize this to $SU(N)$, $N>2$.} 
\begin{eqnarray}
U_\mu(x+N \hat \jmath) = U_\mu^*(x) = \sigma_j U_\mu(x) \sigma_j, \quad
\Phi(x+N \hat \jmath) = \Phi^*(x) = -\sigma_j \Phi(x) \sigma_j
\end{eqnarray}
and $C$-periodic
boundary conditions
\begin{eqnarray}
U_\mu(x+N \hat \jmath) = U_\mu^*(x) = \sigma_2 U_\mu(x) \sigma_2, \quad
\Phi(x+N \hat \jmath) = \Phi^*(x) = -\sigma_2 \Phi(x) \sigma_2
\end{eqnarray}
where $j = 1,2,3$ and $N$ is the number of sites in each of the
spatial directions.
  The point is that the former boundary condition only allows
odd numbers of monopoles, whereas the latter boundary condition only permits even
numbers of monopoles.  In the limit of large inverse temperature $\beta \to
\infty$ (this limit corresponds to extrapolating the temporal extent of the
lattice to infinity), the configurations with the fewest possible monopoles dominate, and
so the mass of the monopole is obtained from
$M = -\lim_{\beta\to\infty} (1/\beta) \ln (Z_\text{tw} / Z_C)$.
In practice one must obtain this quantity as an integral with respect to
some bare lattice parameter.  A scalar mass has been used in previous
studies, and we will continue this practice in our own calculations (directly related
to the parameter $F$ above).  Thus we obtain a finite difference
equation that is to be numerically integrated:
\beq
M(m_{i+1}^2) - M(m_i^2) = -\frac{1}{\beta} \ln \frac{ \left\langle \exp(-(m_{i+1}^2 - m_i^2) \sum_{x} \tr \Phi^2 ) \right\rangle_{m_i^2,\text{tw}} }
{ \left\langle \exp(-(m_{i+1}^2 - m_i^2) \sum_{x} \tr \Phi^2 ) \right\rangle_{m_i^2,C} }
\eeq

The W boson mass is also rather involved.  In this case the difficulty relates
 to Gauss' law on a torus:  we cannot put an isolated charge on a timeslice.  The way that we will circumvent this
is to use the $C$-periodic boundary conditions described above; these project out the zeromode 
of the photon field $A_0$, which would otherwise lead to Gauss' law as a constraint equation when
it is integrated in the path integral.  Additionally, we must form a local gauge transformation
invariant interpolating operator.  This will be done by inserting the W boson operators onto
Polyakov lines that wrap around the lattice:
\beq
C(t) &=& \langle \tr ( {\cal W}_0^-(x) U_0(x+\hat 0) U_0(x + 2 \hat 0) \cdots  U_0(x + (t-1) \hat 0) 
{\cal W}_0^+(x + t \hat 0) U_0(x + (t+1) \hat 0) 
\ddd \cdots U_0(x + (T-1) \hat 0)) \rangle
\eeq
In addition, the operators ${\cal W}_0^-(x)$ must be formed by projections to unitary gauge
based on the local value of the Higgs field $\Phi(x)$.

\section{Variational analysis of scaling dimensions}
In addition to studying the dualities, we also are investigating the
scaling dimensions of operators.  Currently our focus is on the Konishi and
supergravity operators constructed from scalar fields;
even this is nontrivial because the scalars are wrapped up with
the gauge fields in the twisted formulation.
There are essentially two method to access the scalars.  One is to use
$\cU_a(x) \cUb_a(x) - 1 = 2 i B_a(x) + \text{quadratic}$,
based on \myref{Uexpansion}.  The other is to perform a polar
decomposition $\cU_a(x) = H_a(x) U_a(x)$ and then take the
logarithm of the Hermitian matrix, $B_a = \ln H_a(x)$.

\subsection{Konishi operator}
Here we use the interpolating operator
$\Ocal_\text{K.I.} = \sum_{a=1}^5 \tr B_a^2$
where for comparison $B_a$ is constructed by the two separate methods just described.  The $B_a$ fields
are related to the untwisted scalars $\phi_i$, $i=1,\ldots,6$ according to\footnote{Note that here
and in the following, Latin letters from the beginning of the alphabet will have range
$a,b,c=1,\ldots,5$ while Latin letters $i,j,k=1,\ldots,6$.}
$B_a = \sum_{b=1}^5 P_{ab}^{-1} \phi_b$
where $P_{ab}$ are projection operators that relate the twisted and untwisted theories.
Then we find the interpolating operator has an untwisted interpretation of
$\Ocal_\text{K.I.} = \sum_{a,b,c=1}^5 P_{ab}^{-1} P_{ac}^{-1} \tr \phi_b \phi_c = \sum_{a=1}^5 \tr \phi_a^2$,
where the fact that the operators $P$ satisfy $P^T P = 1$ has been used.  Then taking into account
the basic definitions of the Konishi and supergravity operators as irreducible representations
of $SO(6)_R$,
\beq
\Phi^K = \sum_{i=1}^6 \tr \phi_i^2,
\quad \Phi^S_{ij} = \underline{20'}_{ij} = \tr(\phi_i \phi_j) - \delta_{ij} \frac{1}{6} \sum_{k=1}^6 \tr(\phi_k^2)
\eeq
straightforward algebra shows that
$\Ocal_\text{K.I.} = \frac{5}{6} \Phi^K - \Phi^S_{66}$.
Hence when we measure the dimension of the ``Konishi'' operator in the current
approach, what we
are actually getting is a weighted average of the supergravity operator and the Konishi
operator.  It is not straightforward to  get around this in a way
 that is also lattice
gauge invariant.  The trick is that we need the scalar part of the ``gauge field,''
$\phi_6 = \sum_{a=1}^5 P_{5a} A_a$.
The ambiguities, and gauge dependence, have been found
empirically to lead to nonsensical results if we attempt to
build a ``pure'' Konishi operator, which necessarily involves $\phi_6$.
Thus we turn to an alternative approach that is free of these problems.


\subsection{Variational analysis}
There is a linear combination of operators $\Ocal_i(x)$ that we create on the lattice that
has definite scaling dimension $\Delta_\alpha$: 
\beq
\Phi_\alpha(x) = \sum_i d_{\alpha i} \Ocal_i(x)
\label{expand}
\eeq
We define the correlation matrix
$C_{ij}(r) = \vev{ \Ocal_i(x) \Ocal_j(y) }, \; r = ||x-y||$.
If the operators $\Phi_\alpha(x)$ are primary operators of the CFT, then
$\vev{ \Phi_\alpha(x) \Phi_\beta(y) } = \delta_{\alpha \beta} \kappa_\beta r^{-\Delta_\beta}$.
Substituting \myref{expand}, we have
\beq
\sum_{ij} d_{\alpha i} d_{\beta j} C_{ij}(r) 
= \( \frac{r}{r_0} \)^{-\Delta_\beta} \delta_{\alpha \beta} \kappa_\beta r_0^{-\Delta_\beta} 
= \( \frac{r}{r_0} \)^{-\Delta_\beta} \sum_{ij} d_{\alpha i} d_{\beta j} C_{ij}(r_0)
\eeq
Differentiating this equation w.r.t.~$d_{\alpha i}$, we find the generalized eigenvalue problem
\beq
\sum_{j}  C_{ij}(r) d_{\beta j} = \( \frac{r}{r_0} \)^{-\Delta_\beta} \sum_{j}  C_{ij}(r_0) d_{\beta j}
\label{geneig}
\eeq
Note that this only differs from the usual variational analysis in that we have replaced
$e^{-E_n(t-t_0)} \to ( r/r_0)^{-\Delta_\beta}$ 
because we have a CFT with a spectrum of primary operators,\footnote{In this discussion we
are not on the Coulomb branch, but are instead at the superconformal point in moduli space.} 
rather than a gapped theory with
a spectrum of energy eigenvalues $E_n$.  

We take advantage of the fact that $C(r)$ and $C(r_0)$ are symmetric matrices and perform
the Cholesky decomposition
$C(r_0) = Q^T Q$.
Then the generalized eigenvalue problem
can be rewritten as
\beq
(Q^T)^{-1} C(r) Q^{-1} d_\beta = \( \frac{r}{r_0} \)^{-\Delta_\beta} d_\beta
\eeq
The matrix $(Q^T)^{-1} C(r) Q^{-1}$ is symmetric, hence its eigenvalues are
real.  

\section{Outlook}
In forthcoming work,
we will demonstrate
that the lowest lying $\frac{1}{2}-BPS$ states satisfy the tree level relations
in the fully quantum nonperturbative theory.  This will verify a prediction of
S-duality, since the $\frac{1}{2}-BPS$ solitons fill out a multiplet under
$SL(2,\Zbf)$.
The variational approach will be exploited to disentangle the Konishi and supergravity
scaling dimensions (the latter is protected at $\Delta_S=2$, which is an important
check on the lattice theory).  Further down the road,
we will fine tune the lattice to recover the full 16-supercharge supersymmetry
using Monte Carlo renormalization group methods. 

\bibliographystyle{myhunsrt}
\bibliography{lat16}

\begin{thebibliography}{10}

\bibitem{Aharony:2013hda}

\newblock Ofer Aharony, Nathan Seiberg, and Yuji Tachikawa.
\newblock {\em JHEP}, 08:115, 2013, 1305.0318.

\bibitem{Marcus:1995mq}

\newblock Neil Marcus.
\newblock {\em Nucl. Phys.}, B452:331--345, 1995, hep-th/9506002.

\bibitem{Catterall:2007kn}

\newblock Simon Catterall.
\newblock {\em JHEP}, 0801:048, 2008, 0712.2532.

\bibitem{Kaplan:2005ta}

\newblock David~B. Kaplan and Mithat Unsal.
\newblock {\em JHEP}, 0509:042, 2005, hep-lat/0503039.

\bibitem{Rabin:1981qj}

\newblock Jeffrey~M. Rabin.
\newblock {\em Nucl. Phys.}, B201:315--332, 1982.

\bibitem{Becher:1981cb}

\newblock Peter Becher.
\newblock {\em Phys. Lett.}, B104:221--225, 1981.

\bibitem{Becher:1982ud}

\newblock P.~Becher and H.~Joos.
\newblock {\em Z. Phys.}, C15:343, 1982.

\bibitem{Banks:1982iq}

\newblock Tom Banks, Y.~Dothan, and D.~Horn.
\newblock {\em Phys. Lett.}, B117:413--417, 1982.

\bibitem{Catterall:2014vka}

\newblock Simon Catterall, David Schaich, Poul~H. Damgaard, Thomas DeGrand, and
  Joel Giedt.
\newblock {\em Phys. Rev.}, D90(6):065013, 2014, 1405.0644.

\bibitem{Catterall:2011pd}

\newblock Simon Catterall, Eric Dzienkowski, Joel Giedt, Anosh Joseph, and
  Robert Wells.
\newblock {\em JHEP}, 1104:074, 2011, 1102.1725.

\bibitem{Catterall:2014mha}

\newblock Simon Catterall and Joel Giedt.
\newblock {\em JHEP}, 11:050, 2014, 1408.7067.

\bibitem{Catterall:2013roa}

\newblock Simon Catterall, Joel Giedt, and Anosh Joseph.
\newblock {\em JHEP}, 1310:166, 2013, 1306.3891.

\bibitem{Davis:2000kv}

\newblock A.~C. Davis, T.~W.~B. Kibble, A.~Rajantie, and H.~Shanahan.
\newblock {\em JHEP}, 11:010, 2000, hep-lat/0009037.

\bibitem{Davis:2001mg}

\newblock A.~C. Davis, A.~Hart, T.~W.~B. Kibble, and A.~Rajantie.
\newblock {\em Phys. Rev.}, D65:125008, 2002, hep-lat/0110154.

\bibitem{Rajantie:2005hi}

\newblock Arttu Rajantie.
\newblock {\em JHEP}, 01:088, 2006, hep-lat/0512006.

\end{thebibliography}

\end{document}